\documentstyle[sprocl,psfig]{article}

\def\Journal#1#2#3#4{{#1} {\bf #2}, #3 (#4)}

\def\NPB{{\em Nucl. Phys.} B}
\def\PLB{{\em Phys. Lett.}  B}

\newcommand{\mquad}{\!\!\!\!\!\!}
\newcommand{\mqquad}{\!\!\!\!\!\!\!\!\!\!\!\!}

\newcommand{\gl}{\;\;=\;\;}
\newcommand{\tgl}{\!=\!}

\newcommand{\pl}{\;+\;}
\newcommand{\mi}{\;-\;}
\newcommand{\fr}[2]{\frac{#1}{#2}}

\newcommand{\be}{\begin{equation}}
\newcommand{\ee}{\end{equation}}
\newcommand{\bea}{\begin{eqnarray}}
\newcommand{\eea}{\end{eqnarray}}

\newcommand{\bc}{\begin{center}}
\newcommand{\ec}{\end{center}}
\newcommand{\bt}[1]{\begin{tabular}{#1}}
\newcommand{\et}{\end{tabular}}
\renewcommand{\(}{\left(}
\renewcommand{\)}{\right)}
\renewcommand{\[}{\left[}

\newcommand{\eqn}[1]{eq.~(\ref{#1})}

\newcommand{\bm}[1]{\mbox{\mathversion{bold} $\! #1$}}
\newcommand{\bmind}[1]{\mbox{\scriptsize \mathversion{bold} $\! #1$}}

\newcommand{\PH}{\mbox{\mathversion{bold} $\!\Phi$}}

\newcommand{\del}{\partial}

\newcommand{\Tr}{{\rm Tr}\,}

\newcommand{\calL}{{\cal L}}
\newcommand{\calM}{{\cal M}}

\newcommand{\absatz}{{\vspace*{2ex}}}

\begin{document}

{\raggedleft DO-TH 97/16\\
             HD-THEP-97-39\\
             hep-ph/9708317\\[10ex]
}

\title{Nonperturbative correlation masses in the\\ 
3-dimensional SU(2)-Higgs-model}

\author{Andreas Laser}

\address{Fachbereich Physik, Theorie III, Universit\"at Dortmund,\\
D-44221 Dortmund, Germany\\
}

\author{Hans G\"unter Dosch, Michael G.~Schmidt}

\address{Institut  f\"ur Theoretische Physik,
Universit\"at Heidelberg, 
Philosophenweg 16,\\
D-69120 Heidelberg, Germany}

\maketitle\abstracts{
The corelation masses of the 3-dimensional SU(2)-Higgs-model
are calculated from the bound state Green's function. 
For a heavy-light and a light-light system the results 
are reported. We compare with lattice data and find good agreement. 
}
  
The effective action describing the 
high temperature phase of the electroweak 
standard theory is a strongly coupled SU(2)-Higgs-model in three 
dimensions.  
The simple-minded perturbative picture 
would predict long range correlations because of the massless gauge bosons
in this phase,
but the neglect of IR-effects is obviously wrong. 
Lattice calculations confirm indeed that there are confinement 
effects  and that the vector boson acquires a mass.
In a letter\cite{Wir3} we proposed a model for two-dimensional 
bound states of light constituents in 
the hot electroweak phase. The Green's functions of these 
bound states have
been evaluated following a method which was developed by 
Simonov \cite{Simonov3} in 4-dimensional QCD.
In a recent paper \cite{Wir5} we elaborated this model in detail.
The evaluation of the Green's function was generalized and reorganized.
Several modifications of the scalar potential and their influences on the
correlation masses have been discussed. 
The agreement with lattice data was very good. 
In this article we review in short the bound state model and
extend it to a system with a heavy and a light constituent.
We discuss two additional effects on the bound state masses.

\absatz

At high temperatures the standard model can effectively be described by
the three dimensional SU(2)-Higgs-model \cite{KajantieEA1} with
the Lagrangian
\be
\calL \gl
\fr12\,
\Tr\!\bm{F}_{ij} \bm{F}_{ij}\pl
\fr12\,\Tr(\bm{D}_i\bm{\Phi})^{\dagger}(\bm{D}_i\bm{\Phi})\pl
\fr{m_3^2}{2}\,\Tr\!\bm{\Phi}^\dagger\bm{\Phi} \pl
\fr{\lambda_3}{4}\(\Tr\!\bm{\Phi}^\dagger\bm{\Phi}\)^2
\;\;.
\label{L3}
\ee
Even variants of the standard model, 
e.g.\ the minimal supersymmetric standard model, 
can be described by 
this effective theory in a large part of the parameter space \cite{Mikko4}. 
The squared mass $m_3^2$, the quartic coupling $\lambda_3$, 
and  the three dimensional gauge coupling $g_3$ depend on the 
parameters of the fundamental theory and on the temperature.
The quartic coupling has the mass-dimension 1, while $g_3$ has the 
mass-dimension 1/2.

In equation (\ref{L3}) we expressed the gauge field and the scalar field
as $2\times2$-matrices.
This notation uncovers the full 
${\rm SU(2)}_{\rm  gauge}\times {\rm SU(2)_{isospin}}$
invariance of the theory (for details see reference \cite{Wir5}).
The scalar matrix $\bm{\Phi}_{\alpha a}$ has a first gauge index
$\alpha$ and a second isospin index $a$.
\absatz

In the 3-dimensional SU(2)-Higgs-model like in all gauge theories 
in principle everything has to be expressed in terms of gauge invariant
quantities. 
The nonlocal operators corresponding to the bound state of two fundamental
scalars $\PH$ are either
\be\label{nlop}
\Tr\, \PH(x)^\dagger \bm{T}(x,\bar{x}) \PH(\bar{x}) 
\qquad
{\rm or}
\qquad
\Tr\, \PH(x)^\dagger \bm{T}(x,\bar{x}) \PH(\bar{x})\, \tau^i  \quad.
\ee
The matrix $\bm{T}(x,\bar{x})$ with gauge-indices 
is the link-operator defined by
\be\label{linkop}
\bm{T}(x,\bar{x}) \gl {\cal P} \exp\!\( \,- \,i g_3 \int^{\bar{x}}_x 
\bm{A}_k\, dz_k\,\) \quad,
\ee
where ${\cal P}$ denotes the path-ordering operator along some 
given path between $x$ and $\bar x$.
The operator on the left hand side of \eqn{nlop} is an isospin-singlet, 
the operator on the right hand side 
is an isospin-triplet; both are gauge singlets. 

Since they are non-local all possible angular momentum states contribute.
It can be shown that there is no scalar (i.e.\ angular momentum $l\tgl 0$) 
isospin-triplet and no vector (i.e.\ $l\tgl 1$) isospin-singlet.
\absatz

In order to calculate the correlation masses of the bound states 
it is necessary to evaluate the Green's function of the nonlocal
operators (eq.~\ref{nlop}).
The Green's function  $G(x,\bar{x},y,\bar{y})$ connecting 
the two isospin-singlet operators
$\Tr \PH(x)^\dagger \bm{T}(x,\bar{x}) \PH(\bar{x})$ and 
$\Tr \PH(y)^\dagger \bm{T}(y,\bar{y}) \PH(\bar{y})$ 
is built up by two fundamental Green's functions and two link
operators $\bm{T}$. Using the Feynman-Schwinger-representation 
it can be expressed as
\bea\label{gf}
\mqquad&&
G(x,\bar{x},y,\bar{y}) 
\gl
\int_0^\infty \!\!ds \int_0^\infty \!\!d\bar{s} \;
           \exp\( -m_3^2s - \bar{m}_3^2\bar{s} \) \;
             \int\!\! \int \! {\cal D}z\, {\cal D}\bar{z}\,
\\
\mqquad&&
 \times\;\; 
 \exp\!\( -\,\frac{1}{4}\int_0^s\!\! d\tau \,\dot{z}_k^2(\tau) 
 \mi\frac{1}{4}\int_0^{\bar{s}}\!\!d\bar{\tau} \,\dot{\bar{z}}_k^2(\bar{\tau}) 
         \)
        \left\langle {\cal P} \exp\( -ig_3 \oint_{x,\bar{x},\bar{y},y} 
         \mquad\bm{A}_k\, dz_k\) \right\rangle_{\!\!\bmind{A}}  \nonumber
\;\;,
\eea
where $m_3^2$ and $\bar{m}_3^2$ are the squared mass of the 
two fundamental scalars.
$s$ and $\bar s$ are the Schwinger proper time variables. 
$z_k(\tau)$ is a path which connects
$x$ with $y$ and is parameterized by $\tau$; $\tau$ runs from 0 to $s$.
$\dot{z}_k(\tau)$ denotes the derivative of  $z_k(\tau)$
with respect to $\tau$. $\bar z_k(\bar\tau)$ and $\bar\tau$ are
defined accordingly.
The brackets $\langle \ldots \rangle_{\bmind{A}}$ 
symbolize the functional integration over the gauge field $\bm{A}$. 
In principle this averaging contains all perturbative and non-perturbative 
effects.
The last factor is the expectation value of the  Wilson-loop in a
gauge field background. 
The paths  $z(\tau)$ and $\bar z(\bar\tau)$ determine the both sides of
the Wilson-loop.

For a given \,$i\tgl j$\, the Green's function of a triplet is 
identical to the Green's function of a singlet,
but due to the selection rules
the isospin-singlet and isospin-triplet states must have different 
angular momenta.
Therefore they have different masses, as we see below.

The correlation mass of the bound states $\calM$ describes the fall 
off of the Green's function at large distances $\Theta$
\be
\label{DefKorMass}
G(x,\bar{x},y,\bar{y}) \;\;\propto\;\; \exp(\:\!-\,\calM\:\!\Theta\,)
\quad.
\ee
The extension of the states has to be small as compared to their distance.
From equation (\ref{gf}) it follows that 
$\calM$ depends only on $m_3^2$, $\bar m_3^2$  and on the expectation 
value of the Wilson-Loop. 
The other variables are integrated out.
All possible angular momenta contribute to $G(x,\bar{x},y,\bar{y})$.
They have to be isolated later.
\absatz

The correlation masses are calculated by evaluation 
equation (\ref{gf}). 
The crucial assumption in this evaluation is the 
modified area law. The expectation value of the Wilson-Loop
is replaced by a scalar potential $V(r)$ in the exponent. 
This potential can be interpreted as the energy of to static 
scalar fields at distance $r$.

If one has two light scalars with equal mass one has to assume further 
that the classical path dominates the trajectory of the center of mass
of the bound state. If one of the constituent scalars is infinitly heavy,
this is mandatory. 

Using these two assumptions we succeeded in tracing the 
evaluation back to the solution of a two dimensional Schr\"odinger 
equation and a saddle point equation. In the refereces 
\cite{Wir3} and \cite{Wir5} we invetigated the case of
two fundamental scalars of equal mass $\bar{m}_3^2\tgl m_3^2$.
The derivation is, however, easyly modified to the case 
of one infinitly heavy and one light constituent 
(i.e.\ $\bar{m}_3^2 \to \infty$).

Due to the rotational symmetry of the corresponding Hamiltonian 
the Schr\"odinger equation simlifies to the radial equation
\be\label{Radgl}
- \,\frac{1}{2\tilde{\mu}}
\( \fr{\del^2}{\del r^2} \pl \fr1r \fr{\del}{\del r} \mi \fr{l^2}{r^2} 
\) \psi_{nl}(r) \pl
V(r)\; \psi_{nl}(r)
\gl
\epsilon_{nl}(2\tilde{\mu}) \; \psi_{nl}(r)
\;\; .
\ee
Here \,$n\tgl 1,  2, \ldots$\, is the radial and 
\,$l\tgl 0, \pm 1, \pm 2, \ldots$\, is the angular momentum
quantum number. The eigenvalue $\epsilon_{nl}$ depends of course
on the mass-like parameter $\tilde{\mu}$.

In the case of two constituents of equal mass the correlation mass
of the bound state is 
\be
\calM_{nl}(\mu) \gl \frac{m_3^2}{\mu} \pl \mu \pl \epsilon_{nl}(\mu)
\quad,
\ee
with $2\tilde{\mu}\tgl\mu$. In the case of one infinitly heavy scalar 
one finds
\be
\calM_{nl}(\mu) \gl \frac{m_3^2}{2\mu} \pl \fr{\mu}{2} \pl \epsilon_{nl}(2\mu)
\quad,
\ee
with $\tilde{\mu}\tgl\mu$. 
In both cases the parameter $\mu$ is determined by the saddle point
equation
\be\label{sattelpktgl}
\frac{\partial \calM_{nl}(\mu)}{\partial \mu} \gl
0
\quad.
\ee
For a given scalar potential $V(r)$ the correlation masses can 
easyly be calculated with these formulars. 
\absatz

The Green's function \eqn{gf} does not accotunt for the 
scalar self-inter\-ac\-tion. It can be taken into account by adding a 
$\delta$-function-like contribution $V_ {\rm ssi}$ to the scalar potential. 
By comparison with relativistic scatering theory one finds 
\be
V_ {\rm ssi}
\gl
\fr{\lambda_3}{16 m_{\rm constit}^2}\;\delta^2(\vec r)
\quad.
\ee
Here $m_{\rm constit}$ is the mass of one constituent of the bound
state (see reference~\cite{Wir5} for a discussion). 
This correction affects only the s-states. Due to the sing of $V_ {\rm ssi}$ 
the masses of these states are increased. Nevertheless, the effect is 
very small and can be neglected with good accuracy. 

In reference \cite{Simonov4} Dubin e.a.\ calculate a correction which 
compensates the 
overestimated rotational energy of the fluxtube. Adjusting this calculation
to the three dimensional situation one finds that the 
masses of the higher angular momentum states are lowered by:
\be
\label{SimCorrection}
\Delta M_{nl} \gl -\; \frac{16}{3} \;\fr{l^2\,\sigma^2}{M_{nl}^3}
\quad.
\ee
On the other hand we made only two assumptions in calculating 
the correlation masses: the modified area law and the classical 
center of mass trajectory. (For the heavy-light-system only the 
first one is needed.) In addition to these the authors of 
reference \cite{Simonov4} need further assumptions. We therefore can not
see the advantage of their approach and do not take into account
$\Delta M_{nl}$ of equation (\ref{SimCorrection}).
\absatz

\begin{figure}
\unitlength1cm
\begin{picture}(8.0,5.1)
\put(1.3,-0.6){\psfig{figure=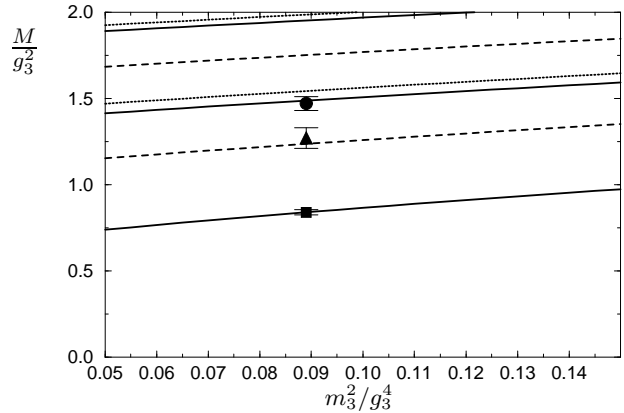,height=2.4in}}
\put(1.5,4.4){\mbox{\large $\fr{M}{g_3^2}$}}
\put(5.7,-0.25){\small $m_3^2/g_3^4$}
\end{picture}
\caption{The spectrum of correlation masses in the light-light-system.
The full lines are the s-masses, the dashed lines are the p-masses
and the dotted lines are the d-masses.}
\label{fig}
\end{figure}

In order to compare with lattice data one has to specify the scalar potential.
We fitted \cite{Wir5} the potentail calculated by 
Ilgenfritz e.a.\ \cite{IlgenfEA1} and calculated the correlation 
masses of the light-light-system with this potential. 
The results are plotted in figure \ref{fig}. 
The intercept of the potential has been adjusted at the 1s-mass
of  Philipsen e.a.\ \cite{PhilipsenEA}, which is plotted as a Square.
The other data points are the 1p-mass (triangle) and 2s-mass (circle)
of reference \cite{PhilipsenEA}. They agree very  well with our predictions. 

In conclusion we find that the bound state model describes 
the dense spectrum of scalar excitations in the 3-dimensional 
SU(2)-Higgs-model very well.


\section*{References}
\begin{thebibliography}{99}



\bibitem{Wir3}         H.G.~Dosch, J.~Kripfganz, A.~Laser and M.G.~Schmidt, 
                         \Journal{\PLB}{365}{213}{1996}.
\bibitem{Simonov3}     Y.A.~Simonov, \Journal{\PLB}{226}{151}{1989}. 
\bibitem{Wir5}         H.G.~Dosch, J.~Kripfganz, A.~Laser and M.G.~Schmidt, 
                         hep-ph/9612450, to appear in {\em Nucl.\ Phys.} B.
\bibitem{KajantieEA1}  K.~Kajantie, M.~Laine, K.~Rummukainen and 
                       M.~Shaposhnikov, \Journal{\NPB}{458}{90}{1996}.
\bibitem{Mikko4}       M.~Laine, \Journal{\NPB}{481}{43}{1996}. 
\bibitem{Simonov4}     A.Yu.~Dubin, A.B.~Kaidalov and 
                         Y.A.~Simonov,  \Journal{\PLB}{323}{41}{1994}.
\bibitem{IlgenfEA1}    E.-M.~Ilgenfritz, J.~Kripfganz, H.~Perlt and A.~Schiller
                       \Journal{\PLB}{356}{561}{1995}.
\bibitem{PhilipsenEA}  O.~Philipsen, M.~Teper and H.~Wittig, 
                        \Journal{\NPB}{469}{445}{1996}.



\end{thebibliography}
\end{document}